\documentclass[10pt, conference]{IEEEtran}
\IEEEoverridecommandlockouts
\usepackage[noadjust]{cite}
\usepackage{amsmath,amssymb,amsfonts}
\usepackage{algorithmic}
\usepackage{graphicx}
\usepackage{textcomp}
\usepackage{xcolor}

\usepackage{comment}
\usepackage{caption}
\usepackage{subcaption}
\usepackage{multirow}   
\usepackage{multicol}   
\usepackage{url}   
\usepackage{comment}
\usepackage{tabularx}
\usepackage{xcolor,colortbl}
\usepackage{hyperref}
\newcolumntype{x}[1]{%
>{\centering\hspace{0pt}}p{#1}}%

\graphicspath{ {./figs/} }

\def\BibTeX{{\rm B\kern-.05em{\sc i\kern-.025em b}\kern-.08em
    T\kern-.1667em\lower.7ex\hbox{E}\kern-.125emX}}
\begin{document}

\title{Graph-Based DDoS Attack Detection in IoT Systems with Lossy Network \\

%\thanks{This material is based upon work supported by Defense Advanced Research Projects Agency (DARPA) under Contract No. HR001120C0160 for the Open, Programmable, Secure 5G (OPS-5G) program. Any views, opinions, and/or findings expressed are those of the author(s) and should not be interpreted as representing the official views or policies of the Department of Defense or the U.S. Government.}
}

\author{\IEEEauthorblockN{Arvin Hekmati}
\IEEEauthorblockA{\textit{Dept. of Computer Science} \\
\textit{University of Southern California}\\
Los Angeles, California, USA \\
hekmati@usc.edu}
\and
\IEEEauthorblockN{Bhaskar Krishnamachari}
\IEEEauthorblockA{\textit{Dept. of ECE and Computer Science} \\
\textit{University of Southern California}\\
Los Angeles, California, USA \\
bkrishna@usc.edu}
}

\maketitle

\begin{abstract} 
This study introduces a robust solution for the detection of Distributed Denial of Service (DDoS) attacks in Internet of Things (IoT) systems, leveraging the capabilities of Graph Convolutional Networks (GCN). By conceptualizing IoT devices as nodes within a graph structure, we present a detection mechanism capable of operating efficiently even in lossy network environments. We introduce various graph topologies for modeling IoT networks and evaluate them for detecting tunable futuristic DDoS attacks. By studying different levels of network connection loss and various attack situations, we demonstrate that the correlation-based hybrid graph structure is effective in spotting DDoS attacks, substantiating its good performance even in lossy network scenarios. The results indicate a remarkable performance of the GCN-based DDoS detection model with an F1 score of up to 91\%. Furthermore, we observe at most a 2\% drop in F1-score in environments with up to 50\% connection loss. The findings from this study highlight the advantages of utilizing GCN for the security of IoT systems which benefit from high detection accuracy while being resilient to connection disruption. 

\end{abstract}

\begin{IEEEkeywords}
IoT, DDoS Attacks, GCN, Lossy Connection
\end{IEEEkeywords}

\section{Introduction}

    The surge in Internet of Things (IoT) device deployment has revolutionized human interaction with their environment, transitioning towards a more integrated physical-digital experience\cite{balaji2019iot}. These devices range from simple household items to complex industrial sensors, aiming for enhanced efficiency and convenience. Predictions by the International Data Corporation (IDC) estimate a rise to 55.9 billion connected IoT devices by 2025, producing 79.4 zettabytes (ZB) of data\cite{reinsel2019iot}. However, this rapid expansion and integration have escalated cybersecurity threats, as indicated by HP Security Research's findings on prevalent vulnerabilities within common IoT devices, including the lack of transport encryption and susceptibility to cross-site scripting~\cite{HPstudy1}. These vulnerabilities underscore the critical need for advanced cybersecurity protocols to ensure the secure growth of IoT networks\cite{hassan2019current}.

    Denial-of-Service (DoS)\cite{liu2009dos} and Distributed Denial-of-Service (DDoS) attacks\cite{nazario2008ddos} represent significant threats within this context. DoS attacks overwhelm targets with excessive requests, impeding legitimate user access\cite{verma2013UDP}, while DDoS attacks employ botnets composed of multiple compromised systems for a synchronized attack\cite{suresh2011ddos}. The IoT's expansive and often minimally secured network forms an ideal platform for DDoS attack propagation.
       
    The Mirai botnet, exploiting IoT vulnerabilities to execute a massive DDoS attack on DNS provider Dyn in 2016, exemplifies the scale of disruption possible, affecting extensive internet operations in the United States\cite{bursztein2017mirai, margolis2017miari}. Similarly, the Reaper botnet, evolved from Mirai, compromised nearly 2.7 million IoT devices\cite{kelley2018reaper}, highlighting the critical need for effective detection and mitigation strategies against DDoS attacks exploiting IoT ecosystems\cite{vishwakarma2020botnet}.
    
    In our previous papers \cite{hekmati2023ton, hekmati2022icccn}, our research centered on the detection of tunable, futuristic DDoS attacks within IoT ecosystems by using correlation-aware neural network models. A critical limitation identified in conventional neural network frameworks such as MLP, CNN, LSTM, Autoencoders, and Transformers is their dependency on fully formed input data. The architectures we introduced necessitate comprehensive node information to accurately predict DDoS attacks emanating from IoT devices. This requirement persists even when a subset of IoT nodes is selected for intercommunication; the neural network models mandate complete input data for effective function. However, within the dynamic and interconnected realm of IoT systems, the potential loss of node correlation information during network transmission poses significant challenges. This scenario underscores the necessity for a mechanism adept at compensating for these missing data points in neural network inputs, a capability that traditional neural networks inherently lack.
    
    In this paper, we delve into the utilization of Graph Convolutional Networks (GCNs) \cite{kipf2016semi} for DDoS attack detection in IoT environments. GCNs emerge as a potent solution, leveraging their robust representation learning capabilities to interpret the complex web of relationships among IoT devices. By conceptualizing IoT devices as nodes within a graph, GCNs inherently accommodate the relational dynamics inherent in IoT networks, thus facilitating the detection process. A standout feature of GCN models lies in their exceptional ability to manage incomplete information, particularly when edges within the graph—representative of the correlation information among IoT nodes—experience disruption or alteration over time. This attribute is especially important in the context of IoT systems, where network connectivity and information flow may be intermittently compromised or altered, leading to potential gaps in correlation data.
    
    The adaptability of GCNs to handle missing edges contrasts sharply with the limitations of classical neural network models, which struggle in the absence of complete input datasets. GCNs, through their graph-based approach, are adept at inferring missing or incomplete relational data, thereby maintaining the integrity of the detection process even when direct correlation information is unavailable. This capability is invaluable in ensuring the resilience of IoT systems against DDoS attacks, particularly in scenarios where attackers may exploit the network's inherent vulnerabilities or disruptions to camouflage their activities. By integrating GCNs into our detection framework, we aim to enhance the robustness and accuracy of DDoS detection in IoT systems, addressing the critical challenges posed by incomplete data and evolving network dynamics.
    
    In our previous study \cite{hekmati2023icmlcn}, we evaluated only the peer-to-peer graph topologies for GCN-based DDoS attack detection. However, in this study, we delve into the construction of various graph topologies, leveraging the network of IoT devices as the foundational framework. We conceptualize IoT devices as nodes within a graph and introduce three distinct methodologies for edge construction to shape the graph's structure. These methodologies encompass: (a) a peer-to-peer approach, where connections are established directly between IoT devices; (b) a network topology-based approach, reflecting the physical or logical structure of the IoT network; and (c) a hybrid approach, which combines elements of both peer-to-peer and network topology strategies. Additionally, we explore various configurations of edge construction, including the consideration of both directed and undirected edges, to accurately model the dynamics of IoT interactions.
    
    Our investigation rigorously evaluates the efficacy of each proposed graph topology in the context of DDoS attack detection within IoT systems. The objective is to ascertain the most effective graph topology for enhancing the accuracy and reliability of DDoS detection mechanisms. Through comprehensive analysis, we aim to identify optimal configurations that not only facilitate accurate modeling of IoT device interactions with the least possible overhead but also significantly improve the detection of DDoS attacks. This evaluation is critical for developing robust defense mechanisms against DDoS attacks, ensuring the security and resilience of IoT networks.
    
    We make the dataset and the attack emulation script, along with our proposed GCN-based detection model, available as an open-source repository online at \url{https://github.com/ANRGUSC/ddos_gcn_paper}.
        
    This paper is structured as follows: Section~\ref{sec:related_works} presents works related to our research field. Subsequently, Section~\ref{sec:system_overview} presents both the attack and detection mechanisms employed in this study, providing a firm grounding in the methodologies utilized. Thereafter, Section~\ref{sec:results} offers a detailed insight into the experimental setup, including an introduction to the dataset leveraged in this research. Furthermore, we present a comprehensive analysis of DDoS attack detection performance by utilizing various graph topologies. The paper concludes with Section~\ref{sec:conclusion}, where we summarize the primary findings of this work and outline promising avenues for future research.

\section{Background and Related Works}
\label{sec:related_works}

    \subsection{Classic ML methods in DDoS Detection}
        Recent advances in machine learning (ML) for DDoS attack detection have significantly contributed to cybersecurity, targeting various network layers and employing a range of ML models. Studies have traditionally concentrated on transport layer attacks, where excessive packet delivery disturb victim server operations. Pioneering works by Doshi et al.~\cite{doshi2018ddosML}, Chen et al.~\cite{chen2020sdnML}, and Mohammed et al.~\cite{mohammed2018sdnML} have introduced classification algorithms such as random forests, K-nearest neighbors, support vector machines, and decision trees, applied within IoT, achieving remarkable accuracies and F1 scores in the range of 0.98 to 0.99.
                
        Explorations beyond the transport layer, by Syed \textit{et al.}~\cite{syed2020brokerML}, extend to application layer DDoS attacks, formulating a DDoS attack detection framework centered around MQTT brokers. They evaluated models including the average one-dependence estimator (AODE), C4.5 decision trees, and MLP, attaining accuracy levels up to 0.99 in virtualized settings.
                
        In the previous papers that studied DDoS attacks, we observed that the IoT nodes' network traffic attributes are significantly different from their benign behavior, making DDoS detection rather easy. In contrast, in our study, we utilize a futuristic DDoS attack with tunable parameter $k$ to carefully control the amount of data sent from IoT devices to the victim server and less expose the attacker. This method aims to mimic regular traffic patterns, thereby concealing the attack's presence. Another standout feature of our proposed model is that it maintains high performance amidst unstable network conditions, such as frequent disconnections, a scenario not extensively covered in existing literature. 

    \subsection{GCN Based Methods in DDoS Detection}
        Graph Convolutional Networks (GCNs) have emerged as a forefront methodology in machine learning, adept at analyzing graph-structured data. By integrating the intrinsic structure of graphs, GCNs excel in uncovering complex patterns and interactions that traditional ML approaches may neglect\cite{zhang2019graph}. Their application extends across various domains, particularly in network security, where the intrinsic connectivity among devices facilitates the detection and mitigation of threats like DDoS attacks. This potential is further investigated in our research. Cao \textit{et al.}~\cite{cao2022gcn} applied Spatial-Temporal Graph Convolutional Networks (ST-GCN) within a Software-Defined Networking (SDN) framework for identifying DDoS activities. Unlike their approach, which was tested against basic DDoS strategies—characterized by distinct differences in attacking and normal traffic patterns—our study adopts a more sophisticated, future-oriented DDoS strategy. This strategy involves adjustable packet volumes, thereby increasing our method's robustness and relevance. Additionally, while their methodology showed limitations in handling unstable, lossy network environments and relied on a centralized router node, posing a single point of failure, our GCN-based approach employs a peer-to-peer topology. This configuration enhances resilience against network instabilities, offering a solution with broader practical application in real-world settings.

\section{System Overview}
\label{sec:system_overview}

    This section presents the construction of both the DDoS attack mechanism leveraging a set of IoT nodes and the GCN-based detection mechanism aiming at securing a victim server by detecting IoT nodes performing DDoS attacks. The scenario of this paper is that we try to detect the source IoT nodes responsible for performing the DDoS attack on a victim server. The following subsections detail each segment individually.

    \subsection{Attack Mechanism}
    \label{sec:attack_mechanism}
        In our study, we adopt the DDoS attack model presented in our preceding work\cite{hekmati2022icccn}, which introduces a novel, adjustable scheme for orchestrating DDoS attacks. This model permits modulation of the attack's intensity via a parameter $k$, which controls the volume distribution of packets sent to the target. Specifically, an increase in $k$ results in a more severe attack, while $k = 0$ mimics the benign behavior of IoT nodes. Throughout the attack duration, all participating IoT nodes are active, employing a variable packet volume distribution governed by the Truncated Cauchy Distribution. This distribution was selected due to its compatibility with historical network traffic analysis efforts\cite{field2002network} and was also observed to be the best distribution for modeling IoT network traffic through our simulations. The parameters for the Truncated Cauchy Distribution-based attack model are defined as follows:
    
        \begin{align}
            \label{eqn:k_x0}
            x_a &= (1+k) \cdot x_b \\
            \label{eqn:k_gamma}
            \gamma_a &= (1+k) \cdot \gamma_b \\
            \label{eqn:k_max}
            m_a &= (1+k) \cdot m_b
        \end{align}
        
        In the specified equations, $x_b$, $\gamma_b$, and $m_b$ represent the location, scale, and maximum packet volume parameters of the benign traffic's truncated Cauchy distribution, respectively. Their counterparts, $x_a$, $\gamma_a$, and $m_a$, denote the corresponding parameters under attack conditions. To ascertain the parameters of the benign traffic's truncated Cauchy distribution, we leveraged the IoT traffic dataset provided by\cite{meidan2018mirai}, which includes real benign traffic data from nine IoT devices. Specifically, we analyzed a 10-second segment of benign packet volumes from the security camera labeled XCS7\_1003 within the dataset\cite{meidan2018mirai}, applying the truncated Cauchy distribution to derive the values of $x_a$, $\gamma_a$, and $m_a$. Although it is feasible to employ three separate parameters to generate new values for $x_a$, $\gamma_a$, and $m_a$, a single, easily adjustable parameter $k$ was chosen for its simplicity.

        Ultimately, the $k$ parameter is designed to modulate the packet volume distribution, with the attack's truncated Cauchy distribution defined via equations (\ref{eqn:k_x0}), (\ref{eqn:k_gamma}), and (\ref{eqn:k_max}). An independent and identically distributed (i.i.d.) sampling method is then utilized to draw from this distribution, determining the packet volumes dispatched at each interval from the IoT nodes towards the target server.
    
        \subsection{Detection Mechanism}
        \label{sec:defense}
            To counteract the DDoS attacks efficiently in IoT systems, we propose graph convolutional networks (GCN) \cite{kipf2016semi} based algorithm as a robust detection mechanism, leveraging a graph representation where the nodes represent IoT devices. This approach, detailed herein, advances from prior papers where we initiated a correlation-aware detection strategy grounded on the concept of sharing the traffic information of IoT nodes with each other, facilitating detecting the futuristic DDoS attacks with low $k$ values. Although correlation-based models proposed in paper \cite{hekmati2023ton} showed promising results in detecting the futuristic DDoS attacks, the method showed vulnerability, predominantly failing in scenarios of network information loss or IoT nodes disconnection. These issues could be addressed by utilizing a GCN-based approach for the detection mechanism.

            Leveraging GCN for DDoS attack detection in IoT systems poses a strong solution due to several reasons. Firstly, GCNs are inherently adept at handling relational data, which is central in network settings, thus providing a natural and efficient framework to capture the complex patterns and dependencies in the network traffic data of IoT devices. Furthermore, it can effectively model and maintain the relational structures even when some nodes are missing, or data is lost, enhancing the robustness of the detection mechanism in volatile network environments typical of many IoT systems. Moreover, GCNs facilitate the extraction of localized and topological features that are pivotal in identifying sophisticated attack patterns, thereby potentially offering a higher detection accuracy compared to traditional machine learning models. Lastly, it allows for the integration of diverse node features (e.g., different node types) in the learning process, thereby enabling a more comprehensive understanding of the underlying patterns and aiding in the timely and accurate detection of DDoS attacks. 
        
            The initial and critical phase in developing a GCN-based DDoS detection model involves the meticulous definition of the graph structure. It is essential to highlight that our scenario is characterized by a temporal dimension, with the graph's features undergoing evolution over time. Consequently, there will be a continuous update of the graph properties, including both the edges and node features, to reflect the dynamic nature of the IoT environment. In the subsequent sections, we will delve into the methodology employed for constructing the graph, detailing the approaches for adapting its structure and attributes to accommodate the temporal variations inherent in our problem space.

        \subsubsection{Defining The Graph Nodes}
        \label{sec:defense-nodes}
            To represent the nodes within our graph, we identify the IoT devices present within the network as the fundamental nodes. Additionally, when incorporating routers into the graph topology, these, too are classified as nodes. The process of defining node attributes involves capturing the network traffic properties of IoT devices, which are subsequently utilized as features for the graph nodes. In scenarios where routers form part of the graph topology, a distinct approach is employed to specify router features. Initially, we focus on routers directly connected to IoT nodes. For a given router $i$, the features of node $i$ are determined by summing the features of IoT devices connected to router $i$ at each time stamp. For routers at higher levels within the graph, this methodology is replicated; the features of a higher-level router are derived by summing the features of directly connected routers, thereby defining the attributes of the higher-level router in a hierarchical manner.

        \subsubsection{Defining The Graph Edges}
        \label{sec:defense-edges}
        
            To accurately reflect the dynamics of IoT systems in our graph model, we propose several mechanisms to define the graph edges. These edges are dynamically established at each time stamp, capturing the network topology, which represents the interconnectivity of nodes and their capacity for mutual traffic information exchange.
            
            \begin{itemize}
                \item Peer-to-Peer: In this strategy, as we can see in figure \ref{fig:gcn-p2p_topology}, each IoT node in the graph is connected to $n$ other IoT nodes. The selection of connections for each IoT node to other nodes within the graph is determined through the following mechanisms:
                \begin{figure}[ht]
                    \centering
                    \includegraphics[width=0.8\columnwidth]{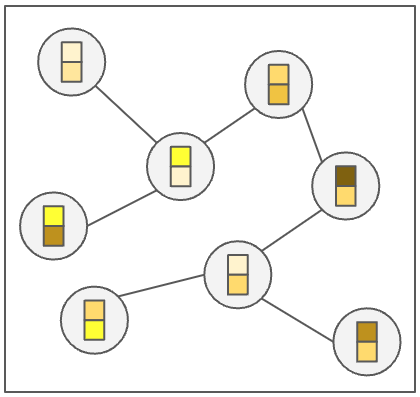}
                    \caption{Peer to peer (p2p) graph topology for GCN-based DDoS attack detection in IoT systems.}
                    \label{fig:gcn-p2p_topology}
                \end{figure}
                
                \begin{itemize}
                    \item \textbf{Distance-based}: In this approach, we create edge connections by aligning each IoT node $i$ with its $n$ geographically closest nodes in the system, thereby leveraging spatial closeness as the defining parameter for connection establishment.
                    
                    \item \textbf{Correlation-based}: In this approach, we connect each IoT node $i$ with $n$ other nodes in the graph that demonstrate the highest Pearson correlation with node $i$ according to their benign activity behavior, advocating a connection grounded on statistical resemblance rather than physical proximity.
                \end{itemize}
            
                \item Network Topology: As presented in figure \ref{fig:gcn-network_topology}, this approach incorporates routers as nodes within the graph, recognizing a hierarchical structure where IoT nodes connect to certain routers, which in turn connect to higher-level routers. This method allows for capturing a graph topology that mirrors the actual network topology, facilitating information flow within the graph.
        
                \begin{figure*}[ht]
                    \centering
                    \includegraphics[width=0.75\textwidth]{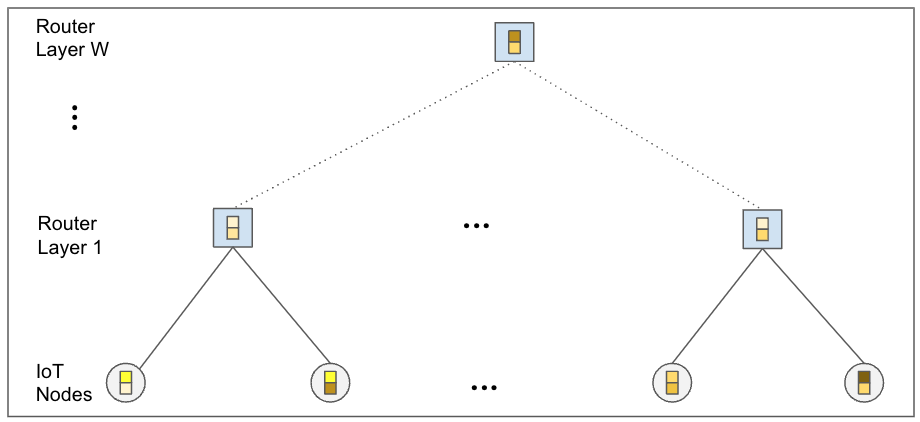}
                    \caption{Network graph topology for GCN-based DDoS attack detection in IoT systems.}
                    \label{fig:gcn-network_topology}
                \end{figure*}

                \item Hybrid: By merging the Peer-to-Peer and Network Topology methods, as presented in figure \ref{fig:gcn-hybrid_topology}, this approach constructs graph edges to create a composite model for edge definition, leveraging both direct device connections and hierarchical network structures.
                \begin{figure}[ht]
                    \centering
                    \includegraphics[width=0.8\columnwidth]{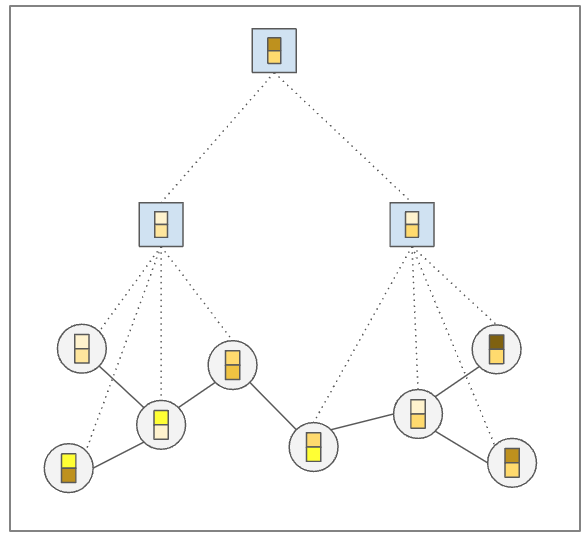}
                    \caption{Hybrid graph topology for GCN-based DDoS attack detection in IoT systems.}
                    \label{fig:gcn-hybrid_topology}
                \end{figure}
        
            \end{itemize}
            
\begin{figure}[ht]
                \centering
                \includegraphics[width=0.8\columnwidth]{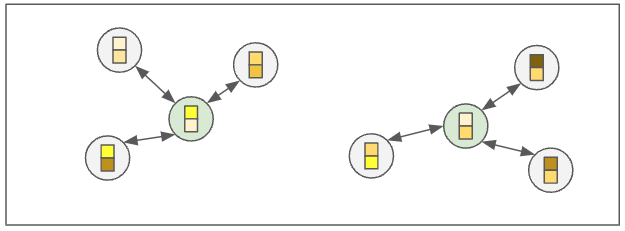}
                \caption{Unidrected graph topology for GCN-based DDoS attack detection in IoT systems.}
                \label{fig:gcn-undirected}
            \end{figure}

            In the process of designing the topology for the graph representation of the IoT system, a critical decision involves determining the connectivity amongst nodes, specifically the orientation of the edges. To this end, we explore the implications of employing both directed and undirected edges within the graph. Figure \ref{fig:gcn-undirected} represents the undirected edges of the graph in which IoT nodes and routers get connected to each other with undirected edges, facilitating information flow in both directions.
        
            When employing directed edges within the graph, a critical consideration is the designation of source and target nodes for each edge. In the context of the graph topologies previously outlined, particularly within the Peer-to-Peer construction method, we explore two distinct scenarios for structuring directed edges:
            
            \begin{itemize}
                \item Node-to-neighbors: In this configuration, as presented in figure \ref{fig:gcn-directed_node_to_neighbors}, we establish connections from each node $i$ to $n$ other nodes within the graph. These connections are based on spatial proximity or the highest correlation to node $i$, with node $i$ acting as the source and the $n$ selected nodes serving as targets. This approach facilitates the directional flow of information from a central node to its neighbors.
                \begin{figure}[ht]
                    \centering
                    \includegraphics[width=0.8\columnwidth]{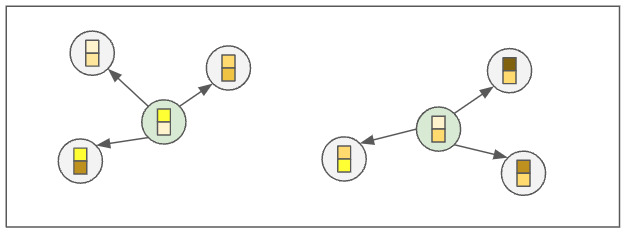}
                    \caption{Directed node-to-neighbors graph topology for GCN-based DDoS attack detection in IoT systems.}
                    \label{fig:gcn-directed_node_to_neighbors}
                \end{figure}
                
                \item Neighbors-to-node: Conversely, as depicted in figure \ref{fig:gcn-directed_neighbors_to_node}, this strategy inverts the direction of information flow by establishing $n$ nodes within the graph as sources and node $i$ as the target. Connections are again determined by either spatial closeness or the correlation to node $i$. This setup models the aggregation of information from multiple nodes towards a single central node.
                \begin{figure}[ht]
                    \centering
                    \includegraphics[width=0.8\columnwidth]{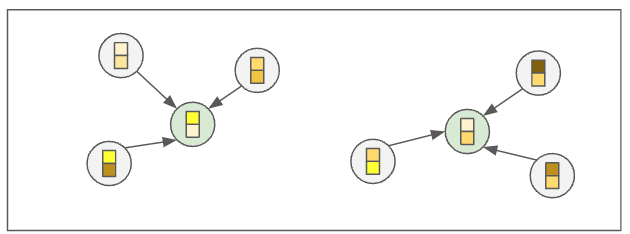}
                    \caption{Directed neighbors-to-node graph topology for GCN-based DDoS attack detection in IoT systems.}
                    \label{fig:gcn-directed_neighbors_to_node}
                \end{figure}
            \end{itemize}
            These scenarios provide a flexible framework for directed edge construction, allowing for tailored information flow models that can be optimized based on the specific dynamics and requirements of the IoT system under investigation. It is important to note that in employing the Network Topology approach, we will only utilize undirected edges. This decision stems from the observation that within the Network Topology framework, using only the directed connection of IoT nodes to routers—or vice versa—does not facilitate the exchange of correlation information among the IoT nodes. Consequently, the utilization of directed edges becomes redundant when adopting the Network Topology method, as it does not align with the primary goal of sharing correlation information across nodes. In scenarios where the Hybrid method is applied, incorporating both routers and IoT nodes within the graph, the preference also shifts towards undirected edges to maintain consistency in information-sharing protocols. An alternative strategy for graph design under the Hybrid method could involve employing undirected edges for router-to-IoT node connections, while reserving directed edges for Peer-to-Peer connections among IoT nodes. However, this particular configuration has not been explored in the current research and is marked for future investigation to assess its viability and effectiveness in enhancing DDoS detection mechanisms within IoT networks.
                
            It is important to note that while undirected edges facilitate broader dissemination of network traffic information among IoT devices, thereby enriching the graph's informational content, such an approach may not be entirely feasible in real-world applications. Specifically, in the deployment of a distributed GCN framework for DDoS detection, the extensive flow of information inherent to a graph populated with undirected edges could lead to substantial network overhead. Consequently, in scenarios characterized by constrained network resources, opting for directed graphs may be advantageous. Directed graphs can effectively streamline the flow of information, preventing potential bottlenecks in the network and thus, maintaining the efficiency and responsiveness of the GCN-based DDoS detection mechanism.
        
        \subsubsection{Model Lossy Connections}
        \label{sec:defense-lossy}
            
\begin{figure*}[ht]
                \centering
                \includegraphics[width=0.95\textwidth]{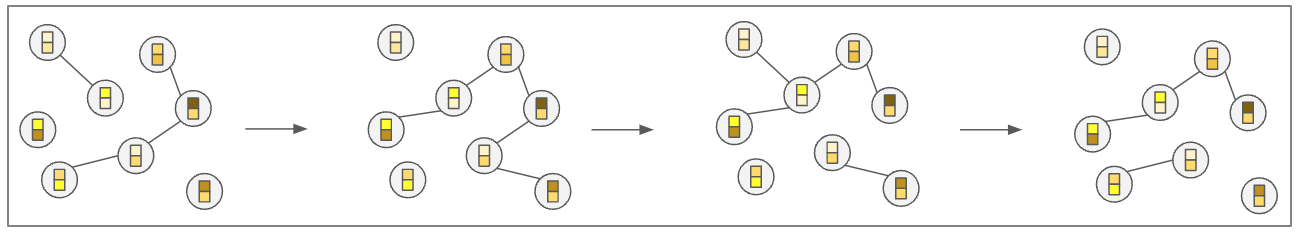}
                \caption{Modeling lossy connections in the graphs for GCN-based DDoS attack detection in IoT systems.}
                \label{fig:gcn-lossy_connection}
            \end{figure*}

            To accurately emulate the phenomenon of connection losses within the network, we introduce a variable, $l$, which specifies the percentage of edges in the graph subject to disconnection at each time stamp as presented in figure \ref{fig:gcn-lossy_connection}. It is important to remember that edges represent the potential for information exchange between IoT nodes and routers. Thus, dropping an edge signifies a failed attempt at information sharing among IoT nodes and routers, due to its loss in the network. This mechanism is crucial for simulating a lossy environment within the IoT system, enabling us to assess the resilience of our proposed GCN model. By incorporating this variable, we aim to evaluate the model's capability to continue detecting DDoS attacks originating from IoT devices despite the challenges posed by intermittent network connectivity.
        
        \subsubsection{Define GCN Model}
        \label{sec:defense-model}
        
            As illustrated in Figure~\ref{fig:gcn-gcn_topology}, the GCN model that we developed encapsulates the process of feature representation learning on graphs. The model operates in three distinct stages:
        
            \begin{figure*}[ht]
                \centering
                \includegraphics[width=0.95\textwidth]{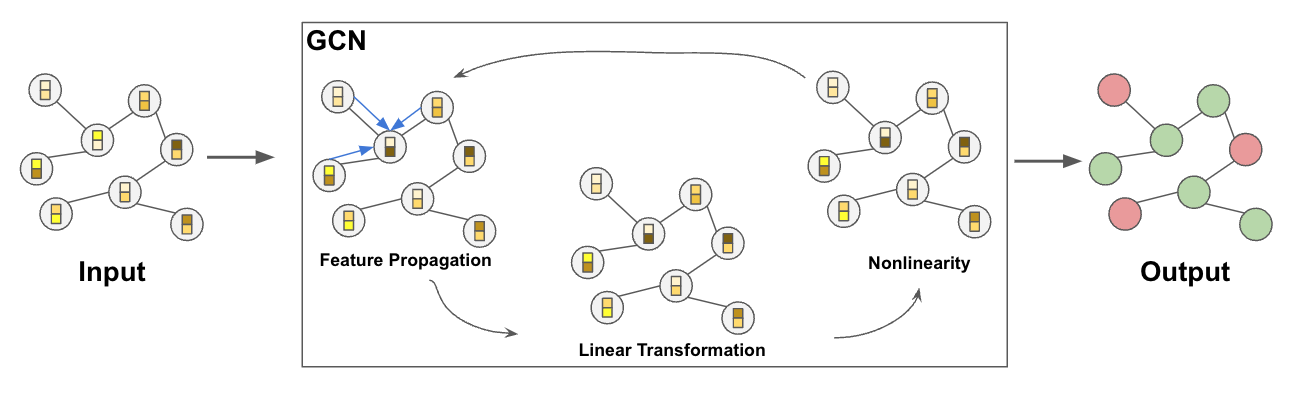}
                \caption{Graph Convolutional Network (GCN) model schematic}
                \label{fig:gcn-gcn_topology}
            \end{figure*}    
        
            \begin{itemize}
                \item \textbf{Input}: Initially, the graph is composed of nodes with their respective feature vectors, represented as squares containing smaller rectangles. These feature vectors serve as the initial state for each node in the graph.
                \item \textbf{Graph Convolutional Operations}:
                \begin{itemize}
                    \item \textit{Feature Propagation}: Here, each node aggregates features from its neighbors, facilitating the diffusion of information across the graph. This step is essential for capturing the graph structure in the feature space.
                    \item \textit{Linear Transformation}: The linear transformation refines the feature representation for the downstream tasks by multiplying the features by the weight matrix.
                    \item \textit{Nonlinearity}: A nonlinearity activation function (usually ReLU) is applied to introduce non-linearity, enabling the model to learn complex patterns from the graph data.
                \end{itemize}
                \item \textbf{Output}: The final output is a graph with nodes that have undergone a transformation, depicted as circles, which represent binary classifications of the IoT nodes predicting which IoT nodes are performing the DDoS attack.
            \end{itemize}

\section{Simulation Results}
\label{sec:results}
In this section, we evaluate the performance of the proposed GCN-based DDoS detection model by utilizing the various topologies discussed in the previous section to find the best configuration for detecting DDoS attacks in IoT systems with lossy networks.

\subsection{Experiment Setup}
\label{sec:results-experiment_setup}
    In the following sections we will discuss the methods we used for designing our experiments.
    
    \subsubsection{Dataset}
    \label{sec:results-dataset}

        In order to evaluate the performance of the proposed GCN model for detecting DDoS attacks in IoT systems with lossy networks, we utilize the dataset derived from real event-driven IoT nodes operative in an urban area, presented in paper \cite{hekmati2023ton}. As discussed previously, this dataset contains several features, including node ID, geographical coordinates (latitude and longitude), time stamps, and packet volumes registered at respective timestamps.
        
        To enrich the dataset, additional features were engineered, presenting the average packet volume transmitted from IoT devices over varying time intervals — 30 minutes, 1 hour, 2 hours, and 4 hours — preceding each timestamp. Furthermore, the packet volume features both in benign and attack states, were synthesized leveraging a truncated Cauchy distribution following the proposed method in section \ref{sec:attack_mechanism}. Table \ref{table:gcn-data_sample} presents some sample data points in the dataset. It is worth mentioning that the number of features used for performing the model training/inference could also be adjusted in the script we provided for the simulations. 

        % Please add the following required packages to your document preamble:
        % \usepackage{graphicx}
        \begin{table*}[ht]
        \centering
        \caption{Sample data points in the training dataset used in GCN-based detection model.}
        \label{table:gcn-data_sample}
        \resizebox{\textwidth}{!}{%
        \begin{tabular}{|c|c|c|c|c|c|c|c|c|c|}
        \hline
        \textbf{NODE} &
          \textbf{LAT} &
          \textbf{LNG} &
          \textbf{TIME} &
          \textbf{ACTIVE} &
          \textbf{PACKET} &
          \textbf{\begin{tabular}[c]{@{}c@{}}PACKET\\ 30 MIN AVG\end{tabular}} &
          \textbf{\begin{tabular}[c]{@{}c@{}}PACKET\\ 1 HR AVG\end{tabular}} &
          \textbf{\begin{tabular}[c]{@{}c@{}}PACKET\\ 2 HR AVG\end{tabular}} &
          \textbf{\begin{tabular}[c]{@{}c@{}}PACKET\\ 4 HR AVG\end{tabular}} \\ \hline
        1 & 10 & 20 & \begin{tabular}[c]{@{}c@{}}2021-01-01\\ 23:00:00\end{tabular} & 0 & 0  & 0    & 0    & 0    & 0    \\ \hline
        1 & 10 & 20 & \begin{tabular}[c]{@{}c@{}}2021-01-01\\ 23:10:00\end{tabular} & 1 & 9  & 3    & 1.5  & 0.75 & 0.38 \\ \hline
        1 & 10 & 20 & \begin{tabular}[c]{@{}c@{}}2021-01-01\\ 23:20:00\end{tabular} & 1 & 11 & 6.67 & 3.33 & 1.67 & 0.83 \\ \hline
        1 & 10 & 20 & \begin{tabular}[c]{@{}c@{}}2021-01-01\\ 23:30:00\end{tabular} & 1 & 10 & 10   & 5    & 2.5  & 1.25 \\ \hline
        \end{tabular}%
        }
        \end{table*}
               
    \subsection{Attack Mechanism Setup}
    \label{subsec:results-attack_mechanism_setup}
    
        The configuration of the DDoS attack mechanism contains an orchestration of attack parameters, including different initiation times at 2 AM, 6 AM, and 12 PM, coupled with varying durations extending for 4, 8, and 16 hours. Furthermore, we also set the IoT nodes that perform the DDoS attack with a participation attack ratio chosen between 0.5 and 1.
        
        A critical element in this setup is the tunable parameter $k$, which dictates the aggression level of the attack, translating to the volume of packets dispatched to the target server. In this experiment, we utilized six distinct values for $k$, ranging from 0 to 1, thereby analyzing attacks mirroring benign traffic (at $k=0$) to those that are far more aggressive (at $k=1$).
        
        In order to prepare a rich training, validation, and testing dataset, we accounted for different attack variations, leveraging diverse combinations of attack properties presented above including initiation times, durations, node participation ratios, and $k$ values.
        
    \subsection{Detection Mechanism Setup}
    \label{subsec:results-detection_mechanism_setup}
    
        We constitute the graphs for the GCN-based detection mechanism based on the methodologies presented in section \ref{sec:defense}. In order to evaluate the Peer-to-Peer method, we connect each node in the graph to four other nodes, i.e., $n=4$, dictated by either spatial proximity or Pearson correlation methods. Furthermore, in the case of using the Network Topology method, we assume the simple scenario of having one router in the network and connecting all IoT nodes to it. It is worth mentioning that although we consider only having one router in the network, the proposed GCN-based defense mechanism utilizing the Network Topology is capable of handling detection in the case of using hierarchical router topology. Additionally, we also evaluate the Hybrid method by constructing the graph using both Peer-to-Peer and Network Topology methods. Finally, for all these combinations, we also evaluate the use of directed or undirected edges, except in the case of having routers in the graph which we will only use undirected edges.
        
        To realistically simulate scenarios of connection losses prevalent in real-world networks, we utilize the $l$ to dictate the percentage of edge drops at individual timestamps, thereby examining network resilience under 0\%, 30\%, and 50\% connection loss scenarios.

        The GCN model used in this paper comprises two GCN convolution layers with 1024 hidden channels. The input dimension to the model is 50 $\times$ 5, which represents 50 nodes in the graph, each node having 5 features which have been described in section \ref{sec:results-dataset}. The node features represent the average packet volume transferred from each node in various time intervals. The edges of the graph do not have any features. The output dimension is 50 $\times$ 1 since we are doing node-level binary classification. During the forward pass, the input graph data undergoes two convolution operations with a ReLU activation function and a dropout layer with a 0.4 probability to help prevent overfitting. The output is obtained via a Sigmoid activation function, making the GCN model compatible with binary cross-entropy-based loss functions for training. We used a customized loss function method for the training which addresses class imbalance in the training dataset by determining class weights from the class frequencies, and then incorporating these weights into a binary cross-entropy loss function. This strategy aims to counteract the influence of class imbalance during training, fostering a model that can more robustly handle imbalanced datasets. The GCN model has been trained for 100 epochs and a batch size of 1024.
            
        Finally, in order to comprehensively analyze the detection performance with a high confidence level in the efficacy of the GCN-based models, we selected ten separate groups of IoT nodes, each containing 50 nodes, to run independent tests, with the subsequent analysis offering 95\% confidence intervals for performance metrics including binary accuracy, F1 score, area under curve (AUC), and recall, thereby providing a comprehensive evaluation of the model's performance.
        
\subsection{Results Discussion}
\label{sec:results_discussion}

    In this section, we analyze the efficacy of the proposed graph convolutional network (GCN) model in detecting DDoS attacks, particularly within the context of lossy network environments. The performance metrics adopted for this examination include binary accuracy, F1 score, area under curve (AUC), and recall, systematically analyzed against variations in the $k$ parameter, which dictates the attack severity by determining the volume of packets dispatched to the victim server during the DDoS attack.
    
    We present the aggregated performance metrics over diverse attack scenarios in each plot. Each plot contains various curves portraying the implications of using various graph topologies discussed in section \ref{sec:defense} and varying degrees of connection loss, represented by parameter $l$. The legends, encoded as ``$X_Y$'', denote the type of Peer-to-Peer topology ($X$) and the corresponding value of connection loss ($Y$), represented by the variable $l$ which indicates the fraction of edges dropped at each time stamp. At most three graph topologies are compared in each plot: 1) Distance, 2)Correlation, and 3) No\_p2p. The value of the connection loss at each time stamp, i.e. $l$, has also been varied between 0, 0.3, and 0.5.
 
\begin{figure}[ht]
        \centering 
        \begin{subfigure}[]{0.24\textwidth}
            \includegraphics[width=\textwidth]{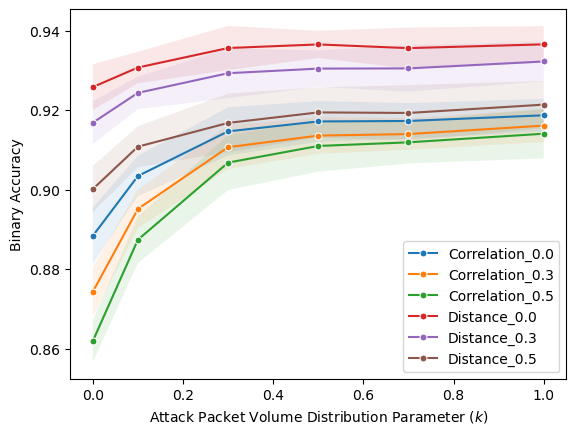}
            \caption{Binary Accuracy}
            \label{fig:gcn-directed_node_to_neighbors_binary_accuracy}
        \end{subfigure}
        \begin{subfigure}[]{0.24\textwidth}
            \includegraphics[width=\textwidth]{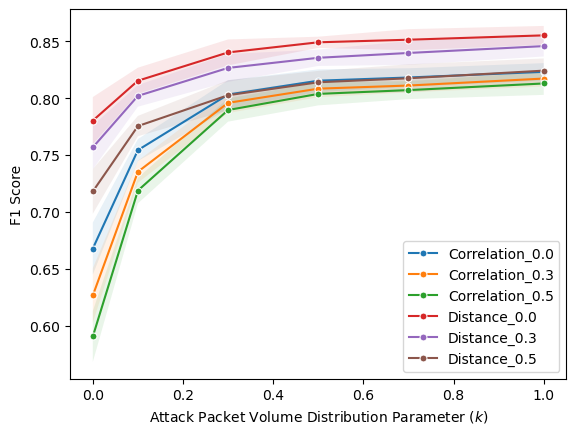}
            \caption{F1 Score}
            \label{fig:gcn-directed_node_to_neighbors_f1_score}
        \end{subfigure}
        \begin{subfigure}[]{0.24\textwidth}
            \includegraphics[width=\textwidth]{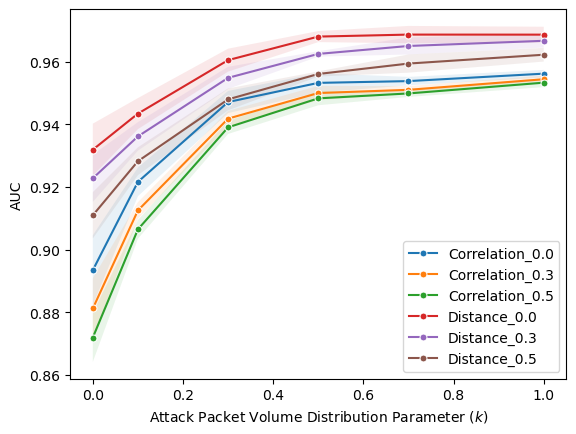}
            \caption{Area Under Curve}
            \label{fig:gcn-directed_node_to_neighbors_auc}
        \end{subfigure}
        \begin{subfigure}[]{0.24\textwidth}
            \includegraphics[width=\textwidth]{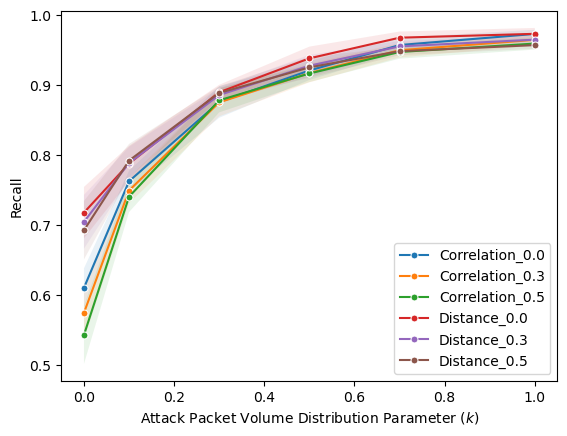}
            \caption{Recall}
            \label{fig:gcn-directed_node_to_neighbors_recall}
        \end{subfigure}
        \caption{Compare the performance of GCN-based DDoS detection model using the Peer-to-Peer topology directed graph with node-to-neighbors edges for graph construction}
        \label{fig:gcn-directed_node_to_neighbors_performance}
    \end{figure}

    Figure \ref{fig:gcn-directed_node_to_neighbors_performance} demonstrates the performance of the Peer-to-Peer topology utilizing directed edges and a node-to-neighbors edge construction approach, across varied attack packet volume distribution parameters ($k$). As depicted in Figure \ref{fig:gcn-directed_node_to_neighbors_f1_score}, for both Distance-based and Correlation-based Peer-to-Peer topologies, the F1 score escalates with the increment of $k$. This phenomenon indicates that higher $k$ values, despite intensifying the attack, concurrently elevate the detectability of the attacker. The Distance-based topology outperforms the Correlation-based approach, where, in scenarios without connection loss (i.e., $l = 0$), the Distance-based methodology achieves an F1 score ranging from 0.78 to 0.85, surpassing the Correlation-based technique, which scores between 0.67 and 0.81. The resilience of the proposed GCN model is evident as the F1 score declines with an increase in connection loss ($l$), with the Distance-based method showing a maximum decrease of 3\% from no connection loss to 50\% connection loss, highlighting its robustness against connection disruptions.

\begin{figure}[ht]
        \centering 
        \begin{subfigure}[]{0.24\textwidth}
            \includegraphics[width=\textwidth]{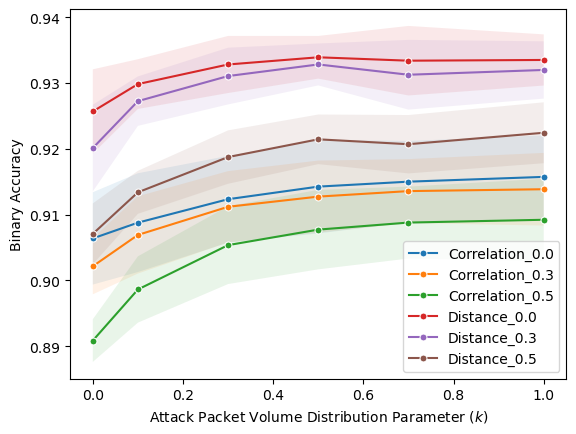}
            \caption{Binary Accuracy}
            \label{fig:gcn-directed_neighbors_to_node_binary_accuracy}
        \end{subfigure}
        \begin{subfigure}[]{0.24\textwidth}
            \includegraphics[width=\textwidth]{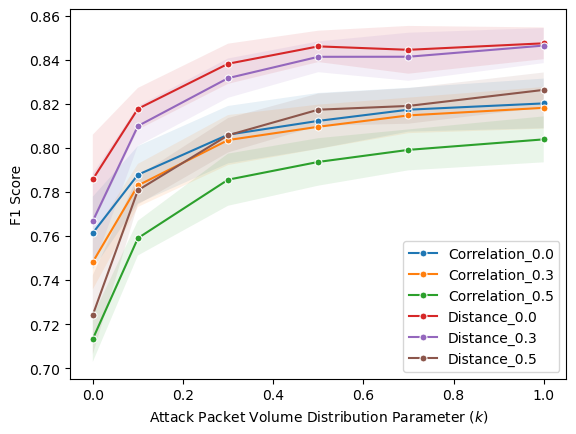}
            \caption{F1 Score}
            \label{fig:gcn-directed_neighbors_to_node_f1_score}
        \end{subfigure}
        \begin{subfigure}[]{0.24\textwidth}
            \includegraphics[width=\textwidth]{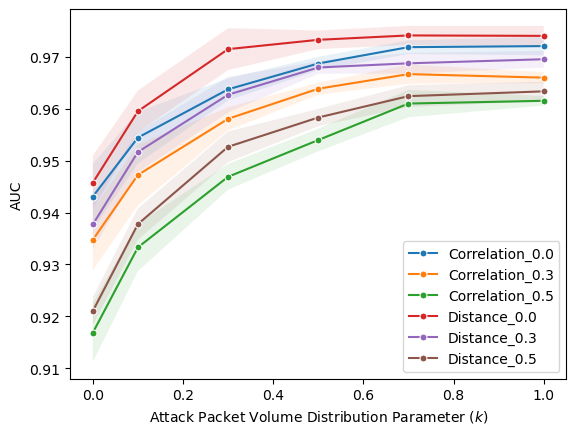}
            \caption{Area Under Curve}
            \label{fig:gcn-directed_neighbors_to_node_auc}
        \end{subfigure}
        \begin{subfigure}[]{0.24\textwidth}
            \includegraphics[width=\textwidth]{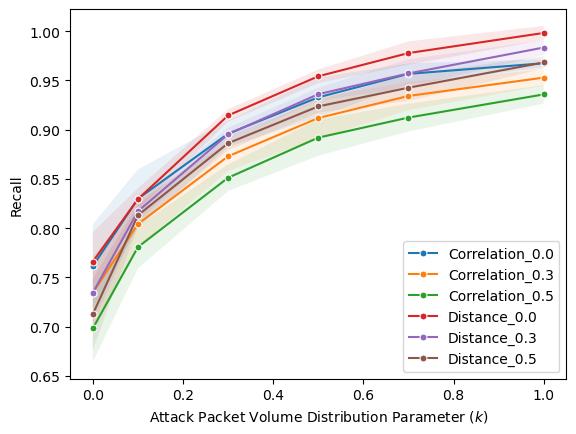}
            \caption{Recall}
            \label{fig:gcn-directed_neighbors_to_node_recall}
        \end{subfigure}
        \caption{Compare the performance of GCN-based DDoS detection model using the Peer-to-Peer topology directed graph with neighbors-to-node edges for graph construction}
        \label{fig:gcn-directed_neighbors_to_node_performance}
    \end{figure}

    Figure \ref{fig:gcn-directed_neighbors_to_node_performance} presents the performance of the Peer-to-Peer topology with directed edges and a neighbors-to-node edge construction strategy under various attack packet volume distribution parameters ($k$). As illustrated in Figure \ref{fig:gcn-directed_neighbors_to_node_f1_score}, for both Distance-based and Correlation-based topologies, the F1 score augments with increasing $k$ values, underscoring the enhanced detectability of more aggressive attacks. The Distance-based topology exhibits superior performance compared to the Correlation-based strategy. Notably, in the absence of connection loss (i.e., $l = 0$), the Distance-based approach records an F1 score from 0.78 to 0.84, marginally higher than the Correlation-based method's 0.76 to 0.81. In terms of the connection loss, the F1 score differential becomes more pronounced with connection loss, where the Distance-based method's F1 score decreases by a maximum of 6\% with 50\% connection loss, reflecting the GCN model's resilience. The directed graph utilizing neighbors-to-node edges showcases an enhanced performance for the Correlation-based algorithm. Especially, for $k=0$, the Correlation-based method achieved an F1 score of 0.67 in the node-to-neighbors method as compared to the 0.76 F1 score that it achieved in the neighbors-to-node method. However, the performance of the Distance-based method remains very similar across both node-to-neighbors and neighbors-to-node configurations. A particularly noteworthy finding is the enhanced resilience of the node-to-neighbors approach to connection losses. Specifically, within the Distance-based method, the degradation in F1 score was limited to a maximum of 3\% when transitioning from a scenario with no connection loss to one with 50\% connection loss, using node-to-neighbors edges. This contrasts with a more pronounced reduction, up to 6\%, observed in the neighbors-to-node edges. In conclusion, when employing directed graphs based on the Peer-to-Peer topology, the Distance-based method not only outperforms alternative approaches in terms of achieving a superior F1 score but also demonstrates great robustness against connection losses, evidenced by a minimal decline in F1 score.

\begin{figure}[ht]
        \centering 
        \begin{subfigure}[]{0.24\textwidth}
            \includegraphics[width=\textwidth]{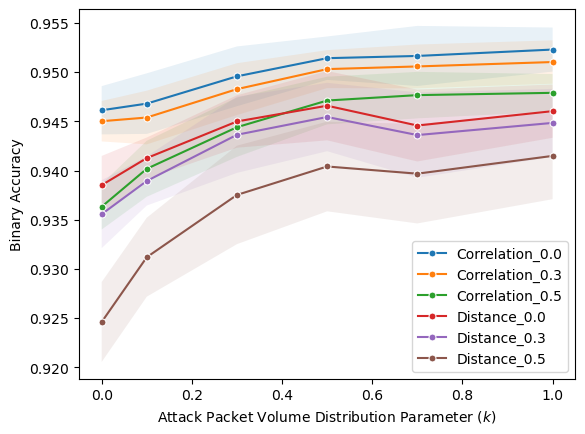}
            \caption{Binary Accuracy}
            \label{fig:gcn-undirected_no_router_binary_accuracy}
        \end{subfigure}
        \begin{subfigure}[]{0.24\textwidth}
            \includegraphics[width=\textwidth]{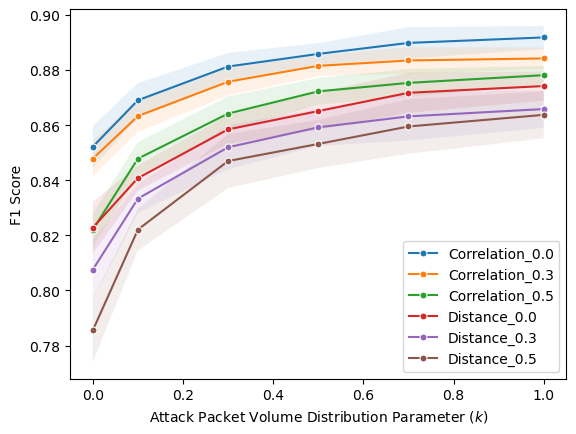}
            \caption{F1 Score}
            \label{fig:gcn-undirected_no_router_f1_score}
        \end{subfigure}
        \begin{subfigure}[]{0.24\textwidth}
            \includegraphics[width=\textwidth]{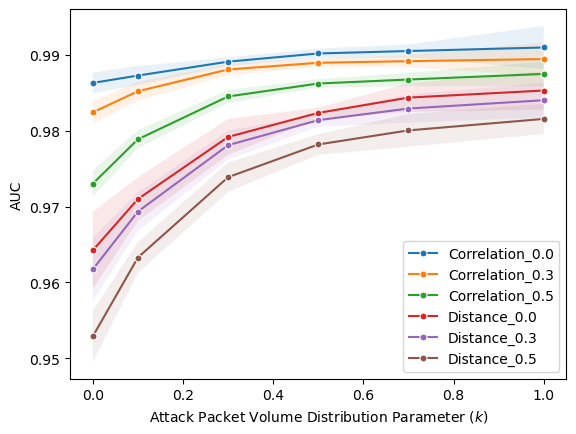}
            \caption{Area Under Curve}
            \label{fig:gcn-undirected_no_router_auc}
        \end{subfigure}
        \begin{subfigure}[]{0.24\textwidth}
            \includegraphics[width=\textwidth]{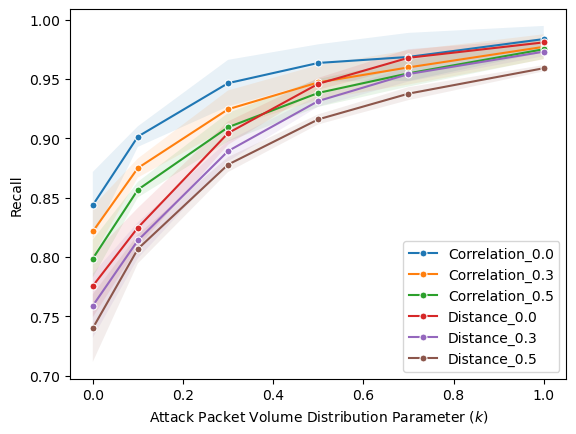}
            \caption{Recall}
            \label{fig:gcn-undirected_no_router_recall}
        \end{subfigure}
        \caption{Compare the performance of GCN-based DDoS detection model using the Peer-to-Peer topology undirected graph for graph construction}
        \label{fig:gcn-undirected_no_router_performance}
    \end{figure}

    Transitioning to undirected graphs, Figure \ref{fig:gcn-undirected_no_router_performance} presents the Peer-to-Peer topology's performance with undirected edges under variable attack packet volume distribution parameters ($k$). As shown in Figure \ref{fig:gcn-undirected_no_router_f1_score}, for both Distance-based and Correlation-based topologies, the F1 score increments with rising $k$ values, affirming that higher aggression levels in attacks improve detectability. Contrary to directed graph scenarios, the Correlation-based topology excels over the Distance-based approach in the undirected configuration, especially when there is no connection loss (i.e., $l = 0$), achieving F1 scores from 0.85 to 0.89 compared to the latter's 0.82 to 0.87. This observation suggests that employing undirected graphs, characterized by enhanced information sharing among IoT nodes, leads to improved performance in DDoS detection. However, it is crucial to acknowledge that while undirected graphs exhibit superior performance, they concurrently increase the network overhead due to the augmented sharing of information between IoT nodes. This elevation in network overhead could potentially result in a network bottleneck, particularly in scenarios where the IoT nodes operate under conditions of limited bandwidth.

\begin{figure}[ht]
        \centering 
        \begin{subfigure}[]{0.24\textwidth}
            \includegraphics[width=\textwidth]{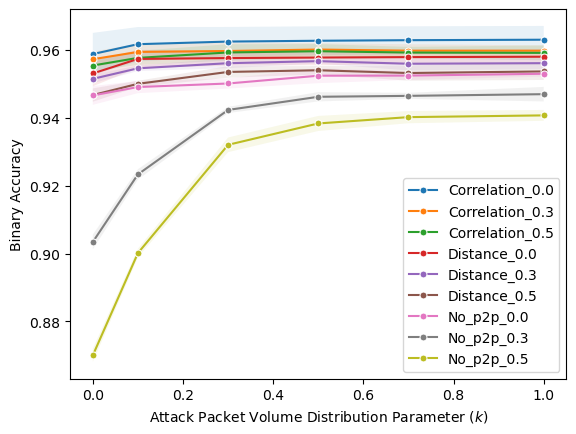}
            \caption{Binary Accuracy}
            \label{fig:gcn-undirected_router_binary_accuracy}
        \end{subfigure}
        \begin{subfigure}[]{0.24\textwidth}
            \includegraphics[width=\textwidth]{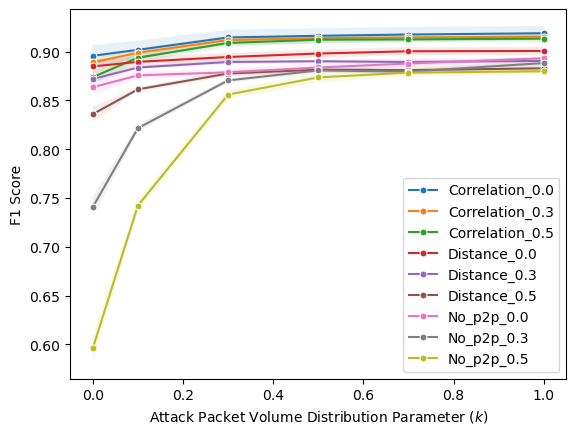}
            \caption{F1 Score}
            \label{fig:gcn-undirected_router_f1_score}
        \end{subfigure}
        \begin{subfigure}[]{0.24\textwidth}
            \includegraphics[width=\textwidth]{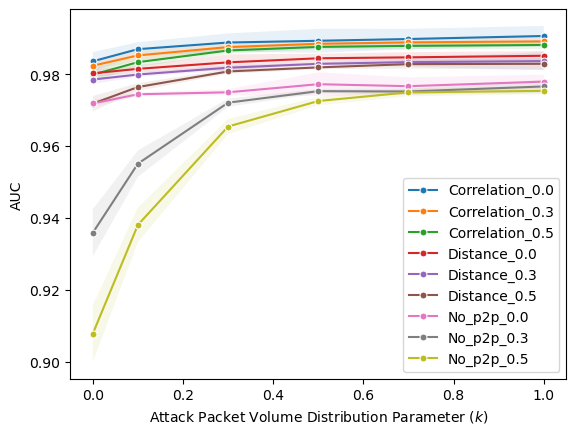}
            \caption{Area Under Curve}
            \label{fig:gcn-undirected_router_auc}
        \end{subfigure}
        \begin{subfigure}[]{0.24\textwidth}
            \includegraphics[width=\textwidth]{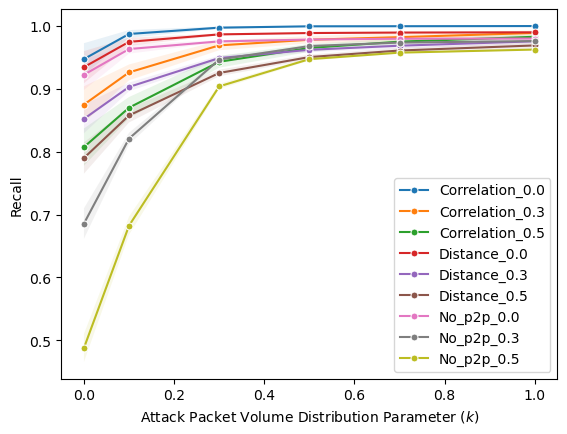}
            \caption{Recall}
            \label{fig:gcn-undirected_router_recall}
        \end{subfigure}
        \caption{Compare the performance of GCN-based DDoS detection model using the Peer-to-Peer topology undirected graph for graph construction}
        \label{fig:gcn-undirected_router_performance}
    \end{figure}

    Lastly, Figure \ref{fig:gcn-undirected_router_performance} depicts the comparative performance of both Network and Hybrid topologies employing the undirected edges approach under various attack packet volume distribution scenarios, parameterized by $k$. The curves that show the Correlation and Distance in the legend are related to the Hybrid method, and the curve that has a No\_p2p legend represents the case of using Network topology without any peer-to-peer connections. As illustrated in Figure \ref{fig:gcn-undirected_router_f1_score}, an increase in $k$ uniformly enhances the F1 score across all evaluated topologies, suggesting that higher $k$ values, despite intensifying the attack, facilitate the attacker's detection. Notably, the Hybrid Correlation-based topology outperforms its counterparts, achieving an F1 score ranging from 0.89 to 0.91 in scenarios without connection loss ($l = 0$), in contrast to the Hybrid Distance-based topology's F1 score, which spans from 0.88 to 0.90. The absence of peer-to-peer connections in the Network topology under similar conditions yields an F1 score between 0.86 and 0.88. A universal decline in F1 score is observed with increasing connection loss ($l$), with the Hybrid Correlation-based topology demonstrating a resilience of up to a 2\% reduction in F1 score from no connection loss to 50\% connection loss. This resilience underscores the proposed GCN model's robustness against connection disruptions. It is interesting to observe that, the exclusion of peer-to-peer connections in the Network topology results in a significant F1 score reduction of approximately 26\% at $k=0$, highlighting the impact of neglecting peer-to-peer edges on DDoS detection, particularly when attackers mimic benign IoT node behavior. Comparative analysis of the GCN model's efficacy with undirected graphs, both with and without a router node, as depicted in Figures \ref{fig:gcn-undirected_router_performance} and \ref{fig:gcn-undirected_no_router_performance}, respectively, reveals superior performance when incorporating a router node within the graph.

\begin{figure}[ht]
        \centering 
        \begin{subfigure}[]{0.24\textwidth}
            \includegraphics[width=\textwidth]{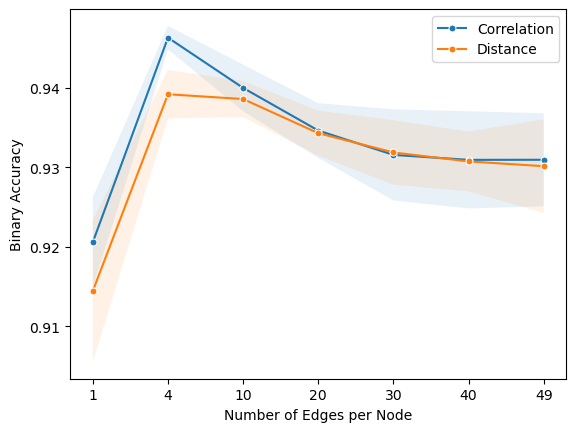}
            \caption{Binary Accuracy}
            \label{fig:gcn-num_edges_per_node_binary_accuracy}
        \end{subfigure}
        \begin{subfigure}[]{0.24\textwidth}
            \includegraphics[width=\textwidth]{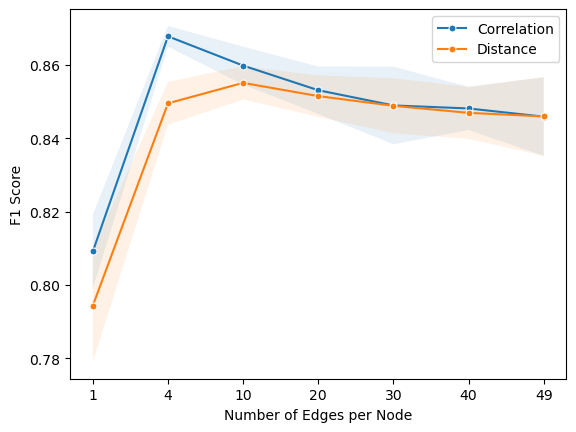}
            \caption{F1 Score}
            \label{fig:gcn-num_edges_per_node_f1_score}
        \end{subfigure}
        \begin{subfigure}[]{0.24\textwidth}
            \includegraphics[width=\textwidth]{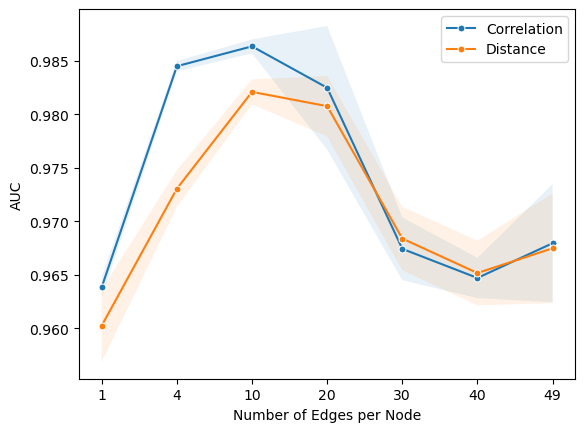}
            \caption{Area Under Curve}
            \label{fig:gcn-num_edges_per_node_auc}
        \end{subfigure}
        \begin{subfigure}[]{0.24\textwidth}
            \includegraphics[width=\textwidth]{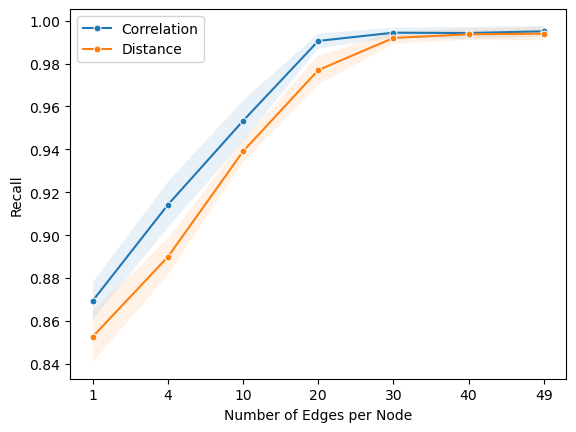}
            \caption{Recall}
            \label{fig:gcn-num_edges_per_node_recall}
        \end{subfigure}
        \caption{Compare the performance of GCN-based DDoS detection model using the Peer-to-Peer topology undirected graph for graph construction and evaluating the number of edges per node.}
        \label{fig:gcn-num_edges_per_node_performance}
    \end{figure}

    Another important factor to evaluate in the GCN-based DDoS detection model is to see how number of edges in the graph would effect the detection performance. In the above results, we evaluated the scenario of having 4 edges per node. In the following, we will consider the cases of having 1 up to 49 edges per node in the graph. Recall that, we have 50 nodes in this simulation results and having 49 edges per node we will essentially have a complete graph. Figure  \ref{fig:gcn-num_edges_per_node_performance} represents the performance of the GCN-based DDoS detection model using the Peer-to-Peer topology undirected graph for graph construction and evaluating the number of edges per node. As we can see in the figure \ref{fig:gcn-num_edges_per_node_f1_score}, as we go from 1 edge per node to 4 edges per node, we see a considerable increase in the F1 score. But after that, the F1 score starts decreasing slightly in the case of having higher number of edges per node. This supports the fact that having too many edges in the graph will results in the scenario of getting smothed features for the nodes and the GCN model would not perform as it is expected to do.

    An essential aspect to analyze within the GCN-based DDoS detection model pertains to the impact of the graph's edge density on detection efficacy. Previously, our analysis was confined to a graph structure incorporating an average of four edges per node. In this simulation, we extend our investigation to encompass graphs varying from one to forty-nine edges per node. Recall that, given the simulation's framework of fifty nodes, attaining forty-nine edges per node results in the formation of a fully connected graph. Figure \ref{fig:gcn-num_edges_per_node_performance} presents the GCN-based DDoS detection model's performance, employing a Peer-to-Peer topology as the foundational structure for the undirected graph, whilst assessing the influence of edge density on node performance. As we can see in Figure \ref{fig:gcn-num_edges_per_node_f1_score}, an initial increase in the F1 score as the edge count per node transitions from one to four, signifying enhanced detection capability. Conversely, a subsequent escalation in edge quantity per node marginally diminishes the F1 score, substantiating the fact that an excessive number of edges fosters feature over-smoothing across nodes, thereby impairing the GCN model's expected performance.

    The main key takeaways from these simulations are the following:

    \begin{itemize}
        \item Across all graph topologies, the performance of the GCN model improves with increasing values of $k$. This observation underscores the principle that higher $k$ values, despite leading to more aggressive attacks, enhance the detectability of the attackers.
        \item In all of the graph topologies, having connection losses results in having a lower detection F1 score.
        \item Among the evaluated topologies, the Hybrid Correlation-based topology with undirected edges emerged as the most effective in detecting DDoS attacks originating from IoT devices, achieving an F1 score range of 0.89 to 0.91. Notably, this topology exhibited the minimal reduction in F1 score due to network connection losses, with a maximum decrease of only 2\% under a 50\% connection loss scenario.
        \item Within directed graph configurations utilizing the Peer-to-Peer topology, the Distance-based method incorporating node-to-neighbor edges achieved the highest F1 score, ranging from 0.78 to 0.85. This is important to note that directed graphs, owing to their reduced network overhead for sharing IoT device correlation information during GCN predictions, are preferable under network constraints affecting IoT devices.
        \item Employing the Network topology without peer-to-peer connections significantly compromises the GCN model's performance in the presence of connection losses. This effect is particularly observed at $k=0$, where a 50\% connection loss led to a dramatic F1 score reduction of approximately 26\%, highlighting the critical importance of peer-to-peer connections in maintaining detection accuracy.
        \item In terms of the edge density in the graph, we observed that having too few or too many edges per node in the graph results in lower performance. In our simulations, we observed that utilizing four edges per node in the case of having 50 nodes in the graph would result in the best performance.
    \end{itemize}

\section{Conclusion}
\label{sec:conclusion}
  
    In this paper, we proposed a GCN-based model for the detection of DDoS attacks within IoT environments. Our investigation highlighted the adaptability of GCNs in managing the complex relational dynamics of IoT devices, conceptualized as nodes within a graph, particularly under conditions of incomplete information. We proposed a robust detection framework that leverages the inherent strengths of GCNs to infer missing or incomplete relational data, thereby maintaining detection integrity even amidst network disruptions.

    We introduced various methodologies for constructing graph topologies, including Peer-to-Peer, Network Topology-based, and Hybrid approaches, to model the IoT system. Our proposed GCN-based detection mechanism, builds upon the premise of sharing traffic information among IoT nodes to facilitate the detection of tunable futuristic DDoS attacks. 
    
    Our experimental setup, utilizing a dataset derived from real event-driven IoT nodes, was designed to simulate a range of DDoS attack scenarios. We examined the effects of different graph construction methodologies, including directed and undirected edges, on the model's detection capabilities. The resilience of our GCN model was put to the test in environments characterized by lossy connections, where we introduced a variable to simulate connection losses within the network.
    
    We extensively analyzed the performance of all proposed graph topologies to find the best configuration for detecting DDoS attacks in IoT systems with lossy networks. We observed that the model's performance improved with higher values of the $k$ parameter, which indicates the volume of packets dispatched during an attack. The Hybrid Correlation-based topology with undirected edges emerged as the most effective configuration, showcasing minimal performance degradation even in scenarios of significant connection loss. Conversely, the exclusive use of a Network topology without peer-to-peer connections markedly diminished the model's efficacy, particularly when the attacker mimicked benign traffic patterns.

    In summary, our findings underscore the potential of GCNs in enhancing DDoS detection within IoT systems, especially in the face of network disruptions and incomplete data. The adaptability of our proposed model to various graph topologies and its resilience against connection losses affirm the viability of GCNs as a cornerstone of future DDoS defense mechanisms in IoT environments.

\section{Acknowledgments}

   This material is based upon work supported by Defense Advanced Research Projects Agency (DARPA) under Contract Number HR001120C0160 for the Open, Programmable, Secure 5G (OPS-5G) program. Any views, opinions, and/or findings expressed are those of the author(s) and should not be interpreted as representing the official views or policies of the Department of Defense or the U.S. Government. This document has been edited with the assistance of ChatGPT. We certify that ChatGPT was not utilized to produce any technical content and we accept full responsibility for the contents of the paper.

\bibliographystyle{IEEEtran}
%\singlespacing
\bibliography{main.bib}

% Generated by IEEEtran.bst, version: 1.14 (2015/08/26)
\begin{thebibliography}{10}
\providecommand{\url}[1]{#1}
\csname url@samestyle\endcsname
\providecommand{\newblock}{\relax}
\providecommand{\bibinfo}[2]{#2}
\providecommand{\BIBentrySTDinterwordspacing}{\spaceskip=0pt\relax}
\providecommand{\BIBentryALTinterwordstretchfactor}{4}
\providecommand{\BIBentryALTinterwordspacing}{\spaceskip=\fontdimen2\font plus
\BIBentryALTinterwordstretchfactor\fontdimen3\font minus
  \fontdimen4\font\relax}
\providecommand{\BIBforeignlanguage}[2]{{%
\expandafter\ifx\csname l@#1\endcsname\relax
\typeout{** WARNING: IEEEtran.bst: No hyphenation pattern has been}%
\typeout{** loaded for the language `#1'. Using the pattern for}%
\typeout{** the default language instead.}%
\else
\language=\csname l@#1\endcsname
\fi
#2}}
\providecommand{\BIBdecl}{\relax}
\BIBdecl

\bibitem{balaji2019iot}
S.~Balaji, K.~Nathani, and R.~Santhakumar, ``Iot technology, applications and
  challenges: a contemporary survey,'' \emph{Wireless personal communications},
  vol. 108, pp. 363--388, 2019.

\bibitem{reinsel2019iot}
D.~Reinsel, ``How you contribute to today’s growing datasphere and its
  enterprise impact,''
  \url{https://blogs.idc.com/2019/11/04/how-you-contribute-to-todays-growing-datasphere-and-its-enterprise-impact/},
  2019, accessed: 09-11-2023.

\bibitem{HPstudy1}
``{Internet of things research study - 2014 report},''
  \url{https://d-russia.ru/wp-content/uploads/2015/10/4AA5-4759ENW.pdf},
  accessed: 11-27-2022.

\bibitem{hassan2019current}
W.~H. Hassan \emph{et~al.}, ``Current research on internet of things ({IoT})
  security: A survey,'' \emph{Computer networks}, vol. 148, pp. 283--294, 2019.

\bibitem{liu2009dos}
W.~Liu, ``Research on dos attack and detection programming,'' in \emph{2009
  Third International Symposium on Intelligent Information Technology
  Application}, vol.~1.\hskip 1em plus 0.5em minus 0.4em\relax IEEE, 2009, pp.
  207--210.

\bibitem{nazario2008ddos}
J.~Nazario, ``Ddos attack evolution,'' \emph{Network Security}, vol. 2008,
  no.~7, pp. 7--10, 2008.

\bibitem{verma2013UDP}
K.~Verma, H.~Hasbullah, and A.~Kumar, ``An efficient defense method against udp
  spoofed flooding traffic of denial of service (dos) attacks in vanet,'' in
  \emph{2013 3rd IEEE International Advance Computing Conference (IACC)}, 2013,
  pp. 550--555.

\bibitem{suresh2011ddos}
M.~Suresh and R.~Anitha, ``Evaluating machine learning algorithms for detecting
  ddos attacks,'' vol. 196, 07 2011, pp. 441--452.

\bibitem{bursztein2017mirai}
E.~Bursztein, ``Inside the infamous mirai iot botnet: A retrospective
  analysis,''
  \url{https://blog.cloudflare.com/inside-mirai-the-infamous-iot-botnet-a-retrospective-analysis/},
  2017, accessed: 09-11-2023.

\bibitem{margolis2017miari}
J.~Margolis, T.~T. Oh, S.~Jadhav, Y.~H. Kim, and J.~N. Kim, ``An in-depth
  analysis of the mirai botnet,'' in \emph{2017 International Conference on
  Software Security and Assurance (ICSSA)}, 2017, pp. 6--12.

\bibitem{kelley2018reaper}
T.~Kelley and E.~Furey, ``Getting prepared for the next botnet attack :
  Detecting algorithmically generated domains in botnet command and control,''
  in \emph{2018 29th Irish Signals and Systems Conference (ISSC)}, 2018, pp.
  1--6.

\bibitem{vishwakarma2020botnet}
\BIBentryALTinterwordspacing
R.~Vishwakarma and A.~K. Jain, ``A survey of ddos attacking techniques and
  defence mechanisms in the iot network,'' \emph{Telecommunication Systems},
  vol.~73, pp. 3--25, 1 2020. [Online]. Available:
  \url{https://link.springer.com/article/10.1007/s11235-019-00599-z}
\BIBentrySTDinterwordspacing

\bibitem{hekmati2023ton}
A.~Hekmati, N.~Jethwa, E.~Grippo, and B.~Krishnamachari, ``Correlation-aware
  neural networks for ddos attack detection in iot systems,'' \emph{arXiv
  preprint arXiv:2302.07982}, 2023.

\bibitem{hekmati2022icccn}
A.~Hekmati, E.~Grippo, and B.~Krishnamachari, ``Neural networks for ddos attack
  detection using an enhanced urban iot dataset,'' in \emph{2022 International
  Conference on Computer Communications and Networks (ICCCN)}.\hskip 1em plus
  0.5em minus 0.4em\relax IEEE, 2022, pp. 1--8.

\bibitem{kipf2016semi}
T.~N. Kipf and M.~Welling, ``Semi-supervised classification with graph
  convolutional networks,'' \emph{arXiv preprint arXiv:1609.02907}, 2016.

\bibitem{hekmati2023icmlcn}
A.~Hekmati and B.~Krishnamachari, ``Graph convolutional networks for ddos
  attack detection in a lossy network,'' \emph{IEEE International Conference on
  Machine Learning for Communication and Networking (IEEE ICMLCN)}, 2024.

\bibitem{doshi2018ddosML}
R.~Doshi, N.~Apthorpe, and N.~Feamster, ``Machine learning ddos detection for
  consumer internet of things devices,'' in \emph{2018 IEEE Security and
  Privacy Workshops (SPW)}, 2018, pp. 29--35.

\bibitem{chen2020sdnML}
Y.-W. Chen, J.-P. Sheu, Y.-C. Kuo, and N.~Van~Cuong, ``Design and
  implementation of iot ddos attacks detection system based on machine
  learning,'' in \emph{2020 European Conference on Networks and Communications
  (EuCNC)}, 2020, pp. 122--127.

\bibitem{mohammed2018sdnML}
S.~S. Mohammed, R.~Hussain, O.~Senko, B.~Bimaganbetov, J.~Lee, F.~Hussain,
  C.~A. Kerrache, E.~Barka, and M.~Z. Alam~Bhuiyan, ``A new machine
  learning-based collaborative ddos mitigation mechanism in software-defined
  network,'' in \emph{2018 14th International Conference on Wireless and Mobile
  Computing, Networking and Communications (WiMob)}, 2018, pp. 1--8.

\bibitem{syed2020brokerML}
\BIBentryALTinterwordspacing
N.~F. Syed, Z.~Baig, A.~Ibrahim, and C.~Valli, ``Denial of service attack
  detection through machine learning for the iot,'' \emph{Journal of
  Information and Telecommunication}, vol.~4, no.~4, pp. 482--503, 2020.
  [Online]. Available: \url{https://doi.org/10.1080/24751839.2020.1767484}
\BIBentrySTDinterwordspacing

\bibitem{zhang2019graph}
S.~Zhang, H.~Tong, J.~Xu, and R.~Maciejewski, ``Graph convolutional networks: a
  comprehensive review,'' \emph{Computational Social Networks}, vol.~6, no.~1,
  pp. 1--23, 2019.

\bibitem{cao2022gcn}
Y.~Cao, H.~Jiang, Y.~Deng, J.~Wu, P.~Zhou, and W.~Luo, ``Detecting and
  mitigating ddos attacks in sdn using spatial-temporal graph convolutional
  network,'' \emph{IEEE Transactions on Dependable and Secure Computing},
  vol.~19, no.~6, pp. 3855--3872, 2022.

\bibitem{field2002network}
T.~Field, U.~Harder, and P.~Harrison, ``Network traffic behaviour in switched
  ethernet systems,'' in \emph{Proceedings. 10th IEEE International Symposium
  on Modeling, Analysis and Simulation of Computer and Telecommunications
  Systems}.\hskip 1em plus 0.5em minus 0.4em\relax IEEE, 2002, pp. 33--42.

\bibitem{meidan2018mirai}
Y.~Meidan, M.~Bohadana, Y.~Mathov, Y.~Mirsky, A.~Shabtai, D.~Breitenbacher, and
  Y.~Elovici, ``N-baiot—network-based detection of {IoT} botnet attacks using
  deep autoencoders,'' \emph{IEEE Pervasive Computing}, vol.~17, no.~3, pp.
  12--22, 2018.

\end{thebibliography}

\end{document}